\RequirePackage{lineno}
\documentclass[twocolumn,showkeywords,superscriptaddress,preprintnumbers,reprint,amsmath,amssymb,epjc,nofootinbib]{revtex4}  
\usepackage{graphicx}
\usepackage{dcolumn}
\usepackage{bm}
\usepackage{float}
\usepackage{color}
\usepackage{appendix}
\usepackage{CJK}
\usepackage[colorlinks,linkcolor=blue,urlcolor=blue,citecolor=blue]{hyperref}
\usepackage{isotope}
\usepackage[normalem]{ulem}
\renewcommand{\sout}{\bgroup \color{red} \ULdepth=-0.5ex \ULset}

\usepackage{amsmath}
\usepackage{cases}
\usepackage{amssymb}

\begin{document}


\title{Hadronization effects on transverse momentum dependent jet fragmentation function in small systems}

\author{Xiang-Pan Duan}
\affiliation{Key Laboratory of Nuclear Physics and Ion-beam Application~(MOE), Institute of Modern Physics, Fudan University, Shanghai $200433$, China}
\affiliation{Shanghai Research Center for Theoretical Nuclear Physics, NSFC and Fudan University, Shanghai $200438$, China}

\author{Wenbin Zhao}
\email{zhaowenb@pku.edu.cn}
\affiliation{Department of Physics and Astronomy, Wayne State University, Detroit, Michigan 48201, USA}

\author{Guo-Liang Ma}
\email{glma@fudan.edu.cn}
\affiliation{Key Laboratory of Nuclear Physics and Ion-beam Application~(MOE), Institute of Modern Physics, Fudan University, Shanghai $200433$, China}
\affiliation{Shanghai Research Center for Theoretical Nuclear Physics, NSFC and Fudan University, Shanghai $200438$, China}


\begin{abstract}
{
 The transverse momentum $j_{\rm T}$-dependent jet fragmentation functions have been investigated in proton+proton (p $+$ p) and  proton+lead (p $+$ Pb) collisions at $\sqrt{s_{\rm NN}} = 5.02~{\rm TeV}$ with a multiphase transport model containing both a simple quark coalescence mechanism and a new hybrid hadronization mechanism with coalescence and fragmentation processes. Hadronized by the new hadronization mechanism, the AMPT model achieves a  quantitative description of the $j_{\rm T}$-dependent jet fragmentation functions measured by ALICE. Besides, no obvious jet-medium interaction and cold nuclear matter effects on the $j_{\rm T}$-dependent jet fragmentation functions in p $+$ Pb collisions were observed. We found the $j_{\rm T}$-dependent jet fragmentation functions are dominated by the quark coalescence contribution for the new hadronization mechanism, which can be decomposed into narrow and wide parts. The root mean square value of the wide part depends on the jet radius $R$ and jet transverse momentum $p_{\rm{T, jet}}$, which is sensitive to different hadronization mechanisms and their components. Therefore, the $j_{\rm T}$-dependent jet fragmentation functions are proposed as a sensitive probe to study the non-perturbative hadronization effect of jets in small colliding systems.
}
\end{abstract}

\maketitle

\section{Introduction}\label{1}

High-energy heavy-ion collisions produce a hot and dense quantum chromodynamics (QCD) matter, called the quark-gluon plasma (QGP)~\cite{Gyulassy:2005NPA750}, which has been observed at the Relativistic Heavy Ion Collider (RHIC)~\cite{BRAHMS:2005NPA757,PHOBOS:2005NPA757,STAR:2005NPA757,PHENIX:2005NPA757,Shen:2020mgh} and the Large Hadron Collider (LHC)~\cite{ATLAS:2010PRL105,ALICE:2011PLB696,CMS:2012EPJC72,Song:2017wtw}. Jets, collimated sprays of particles produced in early-stage hard QCD scattering processes of heavy-ion collisions~\cite{Sterman:1977PRL39,Feynman:1978PRD18,Wang:1997PR280}, loss energy when traversing the hot and dense QGP matter~\cite{Bjorken:1982FERMILAB059,Baier:2000ARNPS50,Qin:2009PRL103}. This phenomenon is referred to as jet quenching~\cite{Gyulassy:1990PLB243,Wang:1992PRL68,Wiedemann:2009sh,Qin:2015IJMPE24,Cao:2021RPP84}, which has been widely used as a hard probe to detect the properties of QGP matter. Not only is substantial evidence of jet quenching observed in experimental measurements~\cite{STAR:2003PRL91,ALICE:2012PRL108,CMS:2014PLB730,ATLAS:2015PRL114}, but also many theoretical calculations has simulated this phenomenon in high-energy heavy-ion collisions ~\cite{Baier:1997NPB484,Wang:2004PLB595,Ma:2013PRC87,Ma:2014PRC89,JET:2014PRC90,Milhano:2016EPJC76,Cao:2018PLB777,Gao:2018PRC97,Chang:2020PLB801,Luo:2022EPJC82,Chen:2019gqo,Wan:2018zpq,Shi:2018izg,Zhang:2017ebx,Xu:2014tda}. 

As one of jet properties, the jet fragmentation function (JFF) describes the momentum distribution of hadrons inside a jet~\cite{Workman:2022ynf,ATLAS:2011EPJC71}. The JFFs can be characterized by the longitudinal momentum fraction $z_{\rm h}$ and the transverse momentum $j_{\rm T}$ carried by the hadrons inside the jet cone. Both longitudinal and transverse momentum JFFs reflect the jet internal structure affected by both the evolution of the QCD parton shower and its hadronization process. The JFFs are only determined phenomenologically due to the non-perturbative nature in principle. It has been argued that the $z_{\rm h}$-dependence is sensitive to the standard collinear fragmentation functions, while the $j_{\rm T}$-dependence probes the transverse momentum-dependent fragmentation functions~\cite{Kang:2017glf,Kang:2019ahe}. For jet quenching, the experimental measurements in high-energy heavy-ion collisions have shown that JFFs are strongly modified in the presence of the QGP, and thus can be used as a sensitive probe for the QCD medium~\cite{CMS:2012nro,CMS:2014PRC90,ATLAS:2014PLB739,ATLAS:2019dsv}. Since many physical effects can affect JFFs, JFFs provide an important tool to study the rich physics associated with jet~\cite{Hwa:2003ic,Majumder:2005jy,Salgado:2004PRL93,Ma:2013PRC88}.

The transverse momentum of the hadron perpendicular to the jet axis, $j_{\rm T}$, gives the transverse structure of the distribution of jet constitute particles. The $j_{\rm T}$-dependent JFFs have been recently measured by the two-particle correlation using the leading-track momentum vector as the axis~\cite{CCOR:1980PLB97,PHENIX:2006PRD74,PHENIX:2006PRC73,ALICE:2019JHEP03} or by jet reconstruction method with the jet axis determined by jet finding algorithm~\cite{CDF:2009PRL102,ATLAS:2011EPJC71,LHCb:2019PRL123,ALICE:2021JHEP09}. The measured $j_{\rm T}$-dependent JFFs have been fitted with a two-component function including a Gaussian function for the narrow part and an inverse Gamma function for the wide part~\cite{ALICE:2019JHEP03,ALICE:2021JHEP09}, which have been compared with Herwig and the PYTHIA 8 and indicates that it is necessary to properly include the $j_{\rm T}$ distribution into the models. In addition, the $j_{\rm T}$-dependent jet fragmentation functions in p $+$ Pb collisions have been also experimentally compared to those in p $+$ p collisions, which shows no significant cold nuclear matter effect in p $+$ Pb collisions. It provides an important baseline for measuring the jet quenching effect on JFFs in relativistic heavy-ion collisions.

The QCD hadronization process is not yet fully understood due to its non-perturbative nature. Fragmentation and quark coalescence are two main well-known phenomenological mechanisms for hadronization, both of which can give good descriptions to experimental measurements of final hadrons, although they work for different kinematics in a complementary way~\cite{Greco:2003xt,Fries:2003vb,Hwa:2003ic}. There have been some efforts to unify the two hadronization mechanisms~\cite{Hwa:2003ic,Majumder:2005jy,Han:2016PRC93}. It has been demonstrated by a ``Hydro-Coal-Frag'' model that the interplay between hydrodynamic freeze-out at low-$p_{\rm T}$, parton coalescence at intermediate $p_{\rm T}$ and string fragmentation at high $p_{\rm T}$ can simultaneously explain the nuclear modification $R_{\rm AA}$, elliptic flow $v_2$ of charged and identified hadrons as well as their flavor dependence in the full range of $p_{\rm T}$ for Pb $+$ Pb collisions at the LHC~\cite{Zhao:2022PRL128}. The same model also provided a good description of measured $p_{\rm T}$ spectra and differential elliptic flow $v_2$($p_{\rm T}$) of identified hadrons over the $p_{\rm T}$ range up to 6 GeV for p $+$ Pb collisions at $\sqrt{s_{\rm NN}} = 5.02~{\rm TeV}$~\cite{Zhao:2020PRL125}. It indicates that the hybrid hadronization works for both large and small colliding systems. In Ref.~\cite{Ma:2013PRC88}, we have studied the jet-medium interaction effect on jet fragmentation function in high-energy heavy-ion collisions based on a multiphase transport (AMPT) model, and show the existence of distinct competition between fragmentation and coalescence for jet hadronization in different $z_{\rm h}$ ranges for different centrality bins. In the present work, we calculate the jet transverse momentum $j_{\rm T}$-dependent fragmentation functions using the jet reconstruction method in p $+$ p and p $+$ Pb collisions at $\sqrt{s_{\rm NN}} = 5.02~{\rm TeV}$ with the AMPT model with a simple quark coalescence hadronization mechanism~\cite{Lin:2005PRC72} and a new hybrid hadronization mechanism containing both dynamical quark coalescence and string fragmentation schemes~\cite{Zhao:2020PRL125,Zhao:2022PRL128}.

The paper is organized as follows. In Sec.~\ref{2}, we introduce the framework of the AMPT model with a jet-triggering technique and two hadronization mechanisms, and the method of jet reconstruction and jet observables. Section.~\ref{3} presents the results on transverse momentum $j_{\rm T}$-dependent jet fragmentation functions and discusses the relevant physics on jet hadronization. A summary is given in Sec.~\ref{4}.

\section{Methodology}\label{2}

\subsection{The AMPT model with old and new hadronization mechanisms}

In this work, we calculate transverse momentum dependent jet fragmentation functions using a string-melting version of AMPT model with two different hadronization mechanisms. The string melting version of the AMPT model contains four main stages of relativistic heavy-ion collisions~\cite{Lin:2005PRC72,Lin:2021mdn}: the initial condition, parton cascade, hadronization, and hadronic rescatterings.

(1) Initial condition. The heavy ion jet interaction generator (HIJING) model~\cite{Wang:1991PRD44,Gyulassy:1994CPC83} simulates the initial condition for p $+$ p, p $+$ A, and A $+$ A collsions.  The transverse density profile of the colliding nuclues is taken to as a Woods-Saxon distribution. The multiple scatterings among incoming  nucleons produce the spatial and momentum distributions of minijet partons and soft excited strings. The nuclear shadowing effect is included via an impact-parameter-dependent but $Q^2$(and flavor)-independent parameterization. A jet triggering technique is used to produce dijet event in the HIJING model, which includes several hard QCD processes with the initial and final state radiations: $g+g\rightarrow g+g$, $g+g\rightarrow q+\bar{q}$, $q+g\rightarrow q+g$, $q+\bar{q}\rightarrow g+g$, $q_{1}+q_{2}\rightarrow q_{1}+q_{2}$, and $q_{1}+\bar{q_{1}}\rightarrow q_{2}+\bar{q_{2}}$ ~\cite{Sjostrand:1994CPC82}. In the string melting mechanism, the primordial hadrons that would be produced from the excited Lund strings, minijet and jet partons in the HIJING model are melted into primordial quarks and antiquarks according to the flavor and spin structures of their valence quarks. The quarks are generated by string melting after a formation time:
\begin{equation}\label{eq.1}
	t_{\rm f}=E_{\rm H}/m^2_{\rm T,H},
\end{equation}
where $E_{\rm H}$ and $m_{\rm T,H}$ represent the energy and transverse mass of the parent hadron. The initial positions of quarks from melted strings are calculated from those of their parent hadrons using straight-line trajectories. Under the initial conditions, a partonic plasma with a dijet is produced.   

(2) Parton cascade. The ZPC model describes parton interactions with two-body elastic scatterings~\cite{Zhang:1998CPC109}. The parton cross section, $\sigma$, is calculated by the leading-order pQCD for gluon-gluon interaction,
\begin{equation}\label{eq.2}
\frac{d \sigma}{d t} = \frac{9 \pi \alpha_{\rm s}^{2}}{2}\left (1+\frac{\mu^{2}}{s}\right) \frac{1}{\left(t-\mu^{2}\right)^{2}},
\end{equation} 
where $\alpha_{\rm s}$ is the strong coupling constant (taken as 0.33), $\mu$ is the Debye screening mass, and $s$ and $t$ are the usual Mandelstam variables. We adjust the Debye mass $\mu$ for different parton interaction cross sections, e.g. 2.265 fm$^{-1}$ is taken for 3 mb and $10^{5}$ fm$^{-1}$ is taken for 0 mb. The parton collisions can not only drive the expansion of the QGP, but also lead to the interactions between jet parton shower and the QGP medium until the freeze out of the whole partonic system. Note that the 0 mb case corresponds to the case switching off parton cascade in which all partons will remain the same as the initial condition. The final results for 0 mb are expected to reflect the combined effect of hadronization and hadronic rescatterings. Therefore, it provides a reference to the 3 mb case, since the difference between them shows the effect of parton cascade. 

(3) Hadronization. In this study, hadronization can be performed by either an old quark coalescence model or a new hybrid hadronization model, respectively. (i) The old quark coalescence model is the original mechanism in the AMPT model~\cite{Lin:2005PRC72}, which simply combines two or three nearest partons into hadrons including mesons and baryons, regardless of the relative momentum among the coalescing partons. The three-momentum is conserved during coalescence. The hadron species are determined by the flavor and invariant mass of combined partons. The simple quark coalescence model includes the formation of all mesons and baryons listed in the HIJING program. The simple coalescence model depends only on the phase space of freezeout partons and does not require the adjustment of free parameters. (ii) The new hybrid hadronization model contains the quark coalescence and string fragmentation. In the quark coalescence process, the momentum distributions of mesons and baryons are defined as~\cite{Han:2016PRC93},
\begin{equation}\label{eq.3}
\begin{split}
\frac{d N_{\rm M}}{d^{3} \mathbf{P}_{\rm M}} = & g_{\rm M} \int d^{3} \mathbf{x}_{1} d^{3} \mathbf{p}_{1} d^{3} \mathbf{x}_{2} d^{3} \mathbf{p}_{2} f_{\rm q}\left(\mathbf{x}_{1}, \mathbf{\rm p}_{1}\right) f_{\bar{\rm q}}\left(\mathbf{x}_{2}, \mathbf{p}_{2}\right)\\
    &\times W_{\rm M}(\mathbf{y}, \mathbf{k}) \delta^{(3)}\left(\mathbf{P}_{\rm M}-\mathbf{p}_{1}-\mathbf{p}_{2}\right),\\
\frac{d N_{\rm B}}{d^{3} \mathbf{P}_{\rm B}} = & g_{\rm B} \int d^{3} \mathbf{x}_{1} d^{3} \mathbf{p}_{1} d^{3} \mathbf{x}_{2} d^{3} \mathbf{p}_{2} d^{3} \mathbf{x}_{3} d^{3} \mathbf{p}_{3} f_{\rm q_{1}}\left(\mathbf{x}_{1}, \mathbf{p}_{1}\right)\\
    &\times f_{\rm q_{2}}\left(\mathbf{x}_{2}, \mathbf{p}_{2}\right) f_{\rm q_{3}}\left(\mathbf{x}_{3}, \mathbf{p}_{3}\right) W_{\rm B}\left(\mathbf{y}_{1}, \mathbf{k}_{1} ; \mathbf{y}_{2}, \mathbf{k}_{2}\right)\\ 
    &\times \delta^{(3)}\left(\mathbf{P}_{\rm B}-\mathbf{p}_{1}-\mathbf{p}_{2}-\mathbf{p}_{3}\right),
\end{split} 
\end{equation}
where $g_{\rm M,B}$ is the statistical factor for forming a meson or baryon of certain spin, $f {\rm _{q,\bar{q}}{(\rm\bf x,p)}}$ are the quark and antiquark phase-space distribution functions and $W_{\rm M,B}$ is the smeared Wigner function of mesons or baryons as a function of the relative coordinates and momenta (${\bf y}$ and ${\bf k}$). Excited states of hadrons are accounted for by the excited states of the harmonic oscillator wave functions. Taking Gaussian wave packets for quarks, the Wigner function of a meson in the $n$-th excited state is given by $W_{\rm M,n}(\bf{y},\bf{k})=\frac{\emph{v}^{\rm n}}{\rm n!}e^{-\emph{v}}$ with $\emph{v}=\frac{1}{2}\left(\frac{{\bf y}^2}{\sigma\rm_M^2}+{\bf k}^2 \sigma\rm_M^2 \right)$, where ${\bf y}$ and ${\bf k}$ are the relative coordinates and momenta between the two constituent quark and antiquark in the meson.  The Wigner function of a baryon in the $n_1$-th and $n_2$-th excited states is given by $W_{\rm B,n_1,n_2} (\mathbf{y}_1 , \mathbf{k}_1 ; \mathbf{y}_2, \mathbf{k}_2)=\frac{\emph{v}_1^{ n_1}}{n_1!}e^{-\emph{v}_1}\cdot\frac{\emph{v}_2^{ n_2}}{n_2!}e^{-\emph{v}_2}$, where $\emph{v}_{\rm i}=\frac{1}{2}\left(\frac{{\bf y}_{\rm i} ^2}{\sigma_{\rm Bi}^2}+{\bf k}_{\rm i} ^2 \sigma_{\rm Bi}^2\right) $  with ${\bf y}_{\rm i}$ and ${\bf k}_{\rm i}$ being the relative coordinates and momenta among the three constituent quarks in the baryon. Following Refs.~\cite{Han:2016PRC93,Zhao:2020PRL125,Zhao:2022PRL128}, the excited meson states up to $n=10$ and excited baryon states up to $n_1+n_2=10$ are included. The width parameters $\sigma_{\rm M}$, $\sigma_{\rm B1}$ and $\sigma_{\rm B2}$ are determined by the radii of formed hadrons~\cite{ParticleDataGroup:2012pjm}. The width parameters of the charged pions, protons, and charged kaons are taken from their empirical charge radii. The same width parameters are used for their isospin partners and their antiparticles as well as their spin resonances. On the other hand, the $\Lambda$ has no charge, then the width parameter is determined instead from the matter radius.  Also, excited states of these hadrons use the same width parameters as their ground states. For more detailed information about the quark coalescence implementation please refer to Ref.~\cite{Han:2016PRC93}. The remnant partons, which can not find the coalescence partners, are merged into the strings in the fragmentation part. Using the ``hadron standalone mode" of PYTHIA 8 ~\cite{Sjostrand:2008CPC178}, these strings will fragment into hadrons. The new hadronization mechanism has been shown to reproduce the $z_{\rm h}$ and $j_{\rm T}$ distributions of the PYTHIA jets that have been hadronized by the default Lund string fragmentation~\cite{Han:2016PRC93}. Recently, the new hybrid hadronization mechanism connected to the coupled linear Boltzmann transport-hydrodynamic model and hadronic afterburner well described the experimental data of $R_{\rm AA}$, $v_2$ and its hadron flavor dependence from low to high $p_{\rm T}$ in Pb $+$ Pb at $\sqrt{s_{NN}}$=5.02 TeV~\cite{Zhao:2022PRL128}.

(4) Hadronic rescatterings. A relativistic transport (ART) model~\cite{Li:1995PRC52} simulates resonance decays and hadronic reactions including both elastic and inelastic scatterings for baryon-baryon, baryon-meson, and meson-meson interactions. The mesons include $\pi$, $\rho$, $\omega$, $\eta$, $K$, $K^*$, and $\phi$ for all possible charges. The baryons and antibaryons include $N$, $\Delta$, $N^*(1440)$, $N^*(1535)$, $\Lambda$,
$\Sigma$, $\Xi$, and $\Omega$. 

In this study, we use the string-melting version of AMPT model with old and new hadronization mechanisms to simulate p $+$ p and p $+$ Pb collisions at $\sqrt{s_{\rm NN}} = 5.02~{\rm TeV}$. We denote the results from old and new hadronization mechanisms as ``AMPT'' and ``New Hadronization'', respectively. The parton cross section is set to $3~\rm mb$, which can effectively both reproduce the collective flow~\cite{Ma:2014pva,Bzdak:2014dia,Bozek:2015swa,He:2015hfa} and  simulate between jet and partonic matter interactions in p $+$ Pb collisions, since the AMPT model with such a parton cross section has successfully described many experimental observables of reconstructed jets, such as $\gamma$-jet imbalance~\cite{Ma:2013PLB724}, dijet asymmetry~\cite{Ma:2013PRC87}, jet fragmentation function~\cite{Ma:2013PRC88,Ma:2014PRC896}, jet shape~\cite{Ma:2014PRC89}, jet anisotropies~\cite{Nie:2014PRC90}, jet transport coefficient~\cite{Zhou:2020EPJA56}, and the redistribution of lost energy~\cite{Gao:2018PRC97,Luo:2022EPJC82}.
In addition, the cross section of $0~\rm mb$ represents the reference without jet-medium interactions. We will focus on the difference of $j_{\rm T}$-dependent JFFs between ``AMPT'' and ``New Hadronization''. Such difference reflects the effect of hadronization process, as they share the same partonic final state and hadronic rescatterings for small colliding systems.

\subsection{The analysis method on $j_{\rm T}$-dependent JFFs}
The transverse momentum of hadrons within a jet, $j_{\rm T}$, is defined as the perpendicular component of the momentum of constituent particle with respect to the reconstructed jet axis,
\begin{equation}\label{eq.4}
j_{\rm T} = \frac{\left|\vec{p}_{\text {jet}} \times \vec{p}_{\text {track}}\right|}{\left|\vec{p}_{\text {jet}}\right|},
\end{equation}
where $\vec{p}_{\text {jet}}$ is the momentum of the jet and $\vec{p}_{\text {track}}$ is the momentum of the tracks. The $j_{\rm T}$-dependent jet fragmentation functions of charged particles are calculated according to,
\begin{equation}\label{eq.5}
D(j_{\rm T}) = \frac{1}{N_{\text {jets}}} \frac{1}{j_{\rm{T, ch}}} \frac{{\rm d} N_{\rm{ch}} }{{\rm d} j_{\rm{T}, \rm{ch}}},
\end{equation}
where $N_{\text {jets}}$ is the number of the reconstructed jets and $N_{\text {ch}}$ is the number of the charged particles inside the reconstructed jet cone. In parallel, we can also define a longitudinal momentum $\xi$-dependent jet fragmentation function, see the Appendix for more details on the results of $\xi$-dependent JFFs. In the ALICE experiment, the $j_{\rm T}$-dependent jet fragmentation function has been fitted by the following two-component function including a Gaussian function and an inverse $\Gamma$ function,
\begin{equation}\label{eq.6}
\frac{1}{N_{\text {jets}}} \frac{{\rm d} N}{j_{\rm{T}, \rm{ch}} {\rm d} j_{\rm{T}, \rm{ch}}}=\frac{B_{2}}{B_{1} \sqrt{2 \pi}} e^{-\frac{j_{\rm{T}}^{2}}{2 B_{1}^{2}}}+\frac{B_{3} B_{5}^{B_{4}}}{\Gamma\left(B_{4}\right)} \frac{e^{-\frac{B_{5}}{j_{\rm{T}}}}}{j_{\rm{T}}^{B_{4}+1}},
\end{equation}
where $B_{\rm i}$ (i=1-5) are the free parameters and $\rm \Gamma$ is the gamma function. The root-mean-square (RMS) value of the $j_{\rm T}$-dependent jet fragmentation function for the narrow component can be obtained from the Gaussian function~\cite{ALICE:2021JHEP09},
\begin{equation}\label{eq.7}
\sqrt{\left\langle j_{\rm{T}}^{2}\right\rangle}=\sqrt{2} B_{1},
\end{equation}
and the RMS value for the wide component can be extracted from the inverse gamma function,
\begin{equation}\label{eq.8}
\sqrt{\left\langle j_{\rm{T}}^{2}\right\rangle}=\frac{B_{5}}{\sqrt{\left(B_{4}-2\right)\left(B_{4}-3\right)}}.
\end{equation}
 
Note that in Ref.~\cite{ALICE:2019JHEP03}, the ALICE experiment found that the final state radiation of jet shower leads to the production of wide component, and defines the remainder other than the final state radiation in the JFF as `hadronization', which is different from our definition of hadronization, by which we mean that hadronization is the process of transition from partons to hadrons.

In this work, the $j_{\rm T}$-dependent jet fragmentation function inside the jet is studied by using the anti-$k_{\rm T}$ algorithm ~\cite{Cacciari:2008JHEP04}, which is implemented using a FastJet framework ~\cite{Cacciari:2012EPJC72}. All particles, containing charged and neutral particles, are considered for the jet reconstruction with a fixed resolution parameter, i.e. jet cone size $R$. However, only charged particles within the jet cones are used to calculate the distributions of jet fragmentation observables. The background for the reconstructed jet is estimated by finding a second cone, which is rotated 90 degrees in $\phi$ from the observed jet cone. The second cone includes all charged particles that can be found within this cone with the same $R$ size as same as the original jet cone, which is applied to calculate the background distribution subtracted from jet fragmentation $j_{\rm T}$-dependent jet fragmentation function. We use the same kinematic cuts as the ALICE experiment~\cite{ALICE:2021JHEP09}, where the transverse momenta of particles need to exceed $0.15~{\rm GeV}/c$ and the jet pseudorapidity satisfy $\left|\eta_{\rm{jet}}\right| < 0.25$.

\section{Results and discussion}\label{3}

\subsection{The $p_{\rm T}$ spectra}\label{3.1}

\begin{figure}[htbp]
\centering
\includegraphics
[width=9cm]{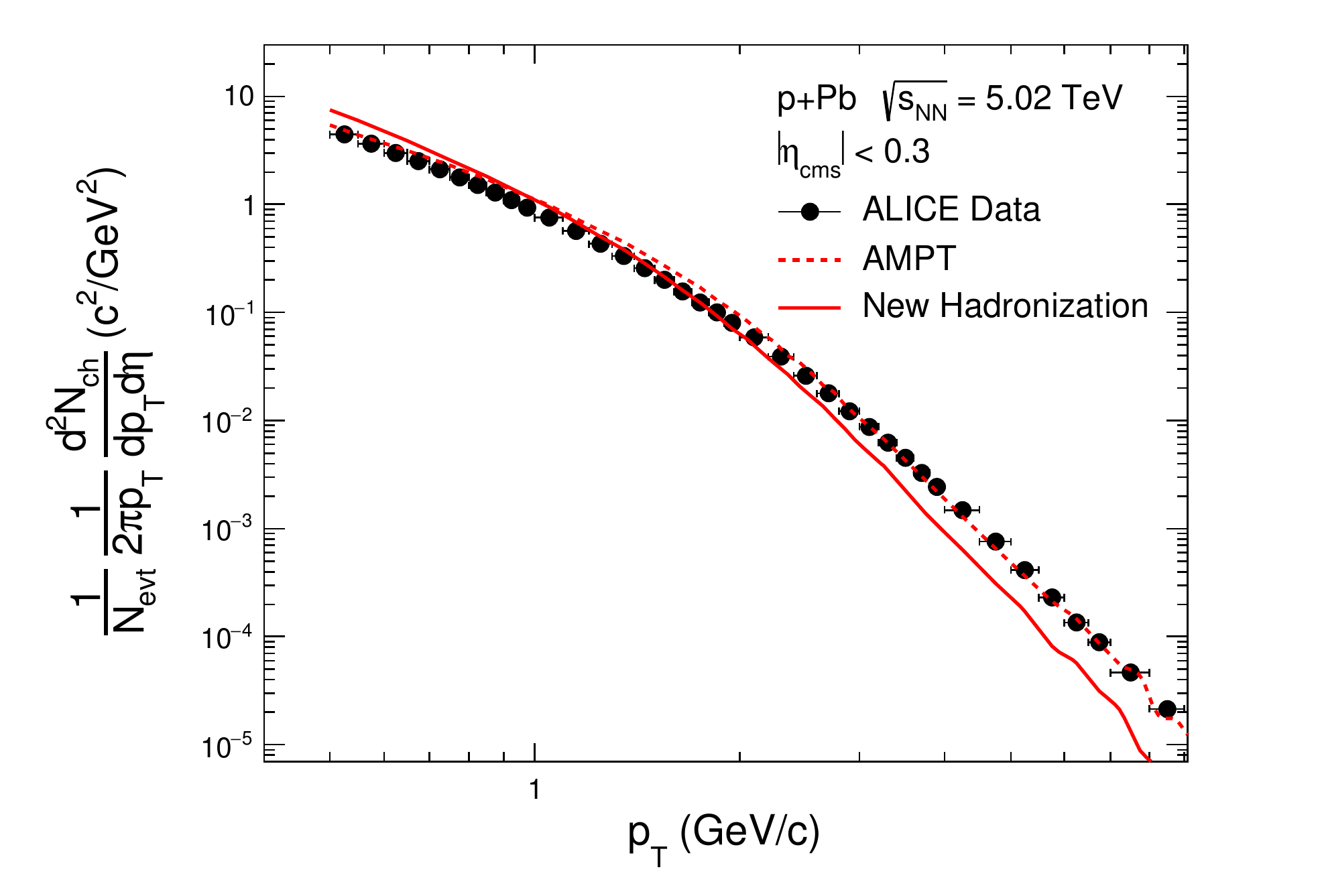}
\caption{The $p_{\rm T}$ spectra of charged particles from the AMPT model with old (dashed curve) and new  (solid curve) hadronization mechanisms in p $+$ Pb collisions, compared to the ALICE data~\cite{ALICE:2012PRL110}.}
\label{fig.1}
\end{figure}

We use two hadronization models, the old simple coalescence model and the new hybrid hadronization model, to calculate the $p_{\rm T}$ spectra of charged particles with $p_{\rm T} > 0.4~{\rm GeV}$/$c$ and $\left|\eta _{\rm cms}\right| < 0.3$ in p $+$ Pb collisions at $\sqrt{s_{\rm NN}} = 5.02~{\rm TeV}$ and compare them with the ALICE data~\cite{ALICE:2012PRL110}. In Fig~\ref{fig.1}, we find that the $p_{\rm T}$ spectra of charged particles can be reproduced better by the old hadronization model than the new hadronization model, as the new hadronization model produces more soft particles in p $+$ Pb collisions.

\begin{figure*}[!htbp]
\centering
\includegraphics
[width=18cm]{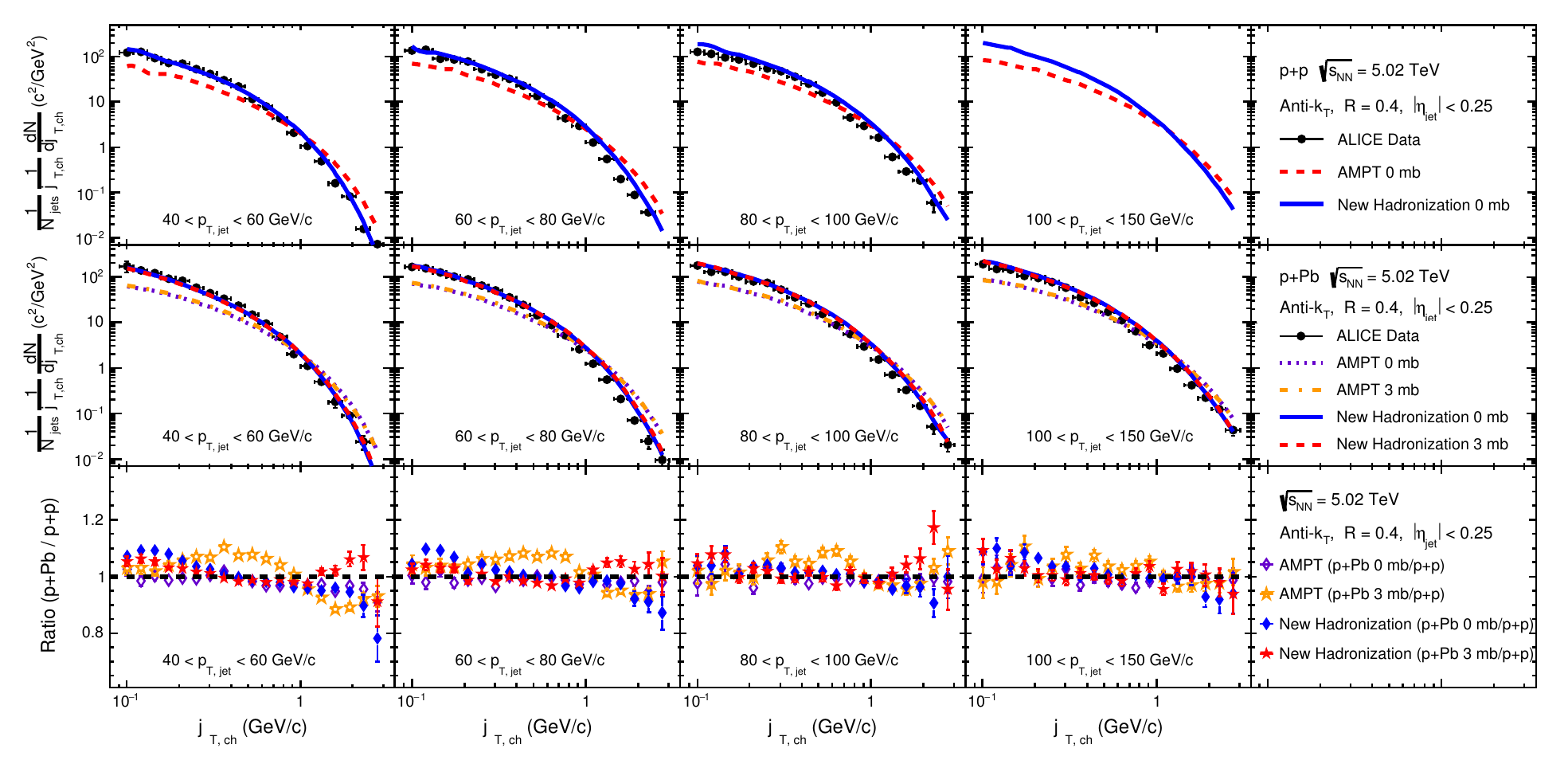}
\caption{Top panels: The $j_{\rm T}$-dependent jet fragmentation functions of charged particles for four $p_{\rm{T, jet}}$ ranges in p $+$ p collisions at $\sqrt{s_{\rm NN}} = 5.02~{\rm TeV}$ from the AMPT model with old and new hadronization mechanisms, compared to the ALICE data~\cite{ALICE:2021JHEP09}. Middle panels: Same as the top panels but for p $+$ Pb collisions at $\sqrt{s_{\rm NN}} = 5.02~{\rm TeV}$. Bottom panels: The ratios of the $j_{\rm T}$-dependent jet fragmentation function in p $+$ Pb collisions (0 or 3 mb) to that in p $+$ p collisions (0 mb) for the AMPT model with the old and new hadronization mechanisms. }
\label{fig.2}
\end{figure*}

\begin{figure}[htb]
\centering
\includegraphics
[width=9cm]{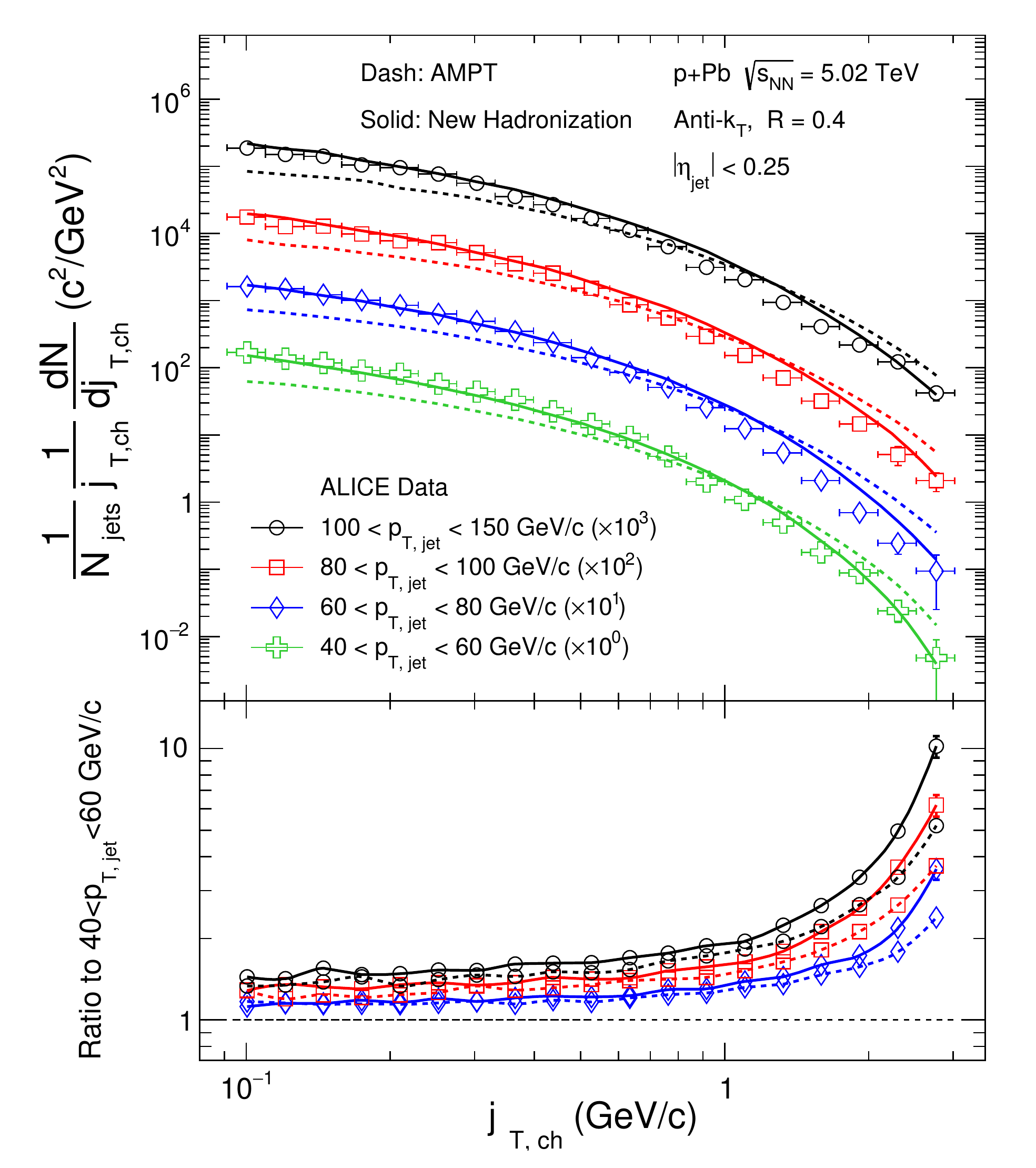}
\caption{Upper panel: The $j_{\rm T}$-dependent jet fragmentation functions of charged particles for four $p_{\rm{T, jet}}$ ranges in p $+$ Pb collisions at $\sqrt{s_{\rm NN}} = 5.02~{\rm TeV}$ from the AMPT model with old (dashed curve) and new hadronization mechanisms (solid curve), compared to the ALICE data~\cite{ALICE:2021JHEP09}. Bottom panel: The ratios of the $j_{\rm T}$-dependent jet fragmentation function for the high $p_{\rm{T, jet}}$ range to that for  40 $<p_{\rm{T, jet}}<$ $60~{\rm GeV}$/$c$.}
\label{fig.3}
\end{figure}

\begin{figure}[htb]
\centering
\includegraphics
[width=9cm]{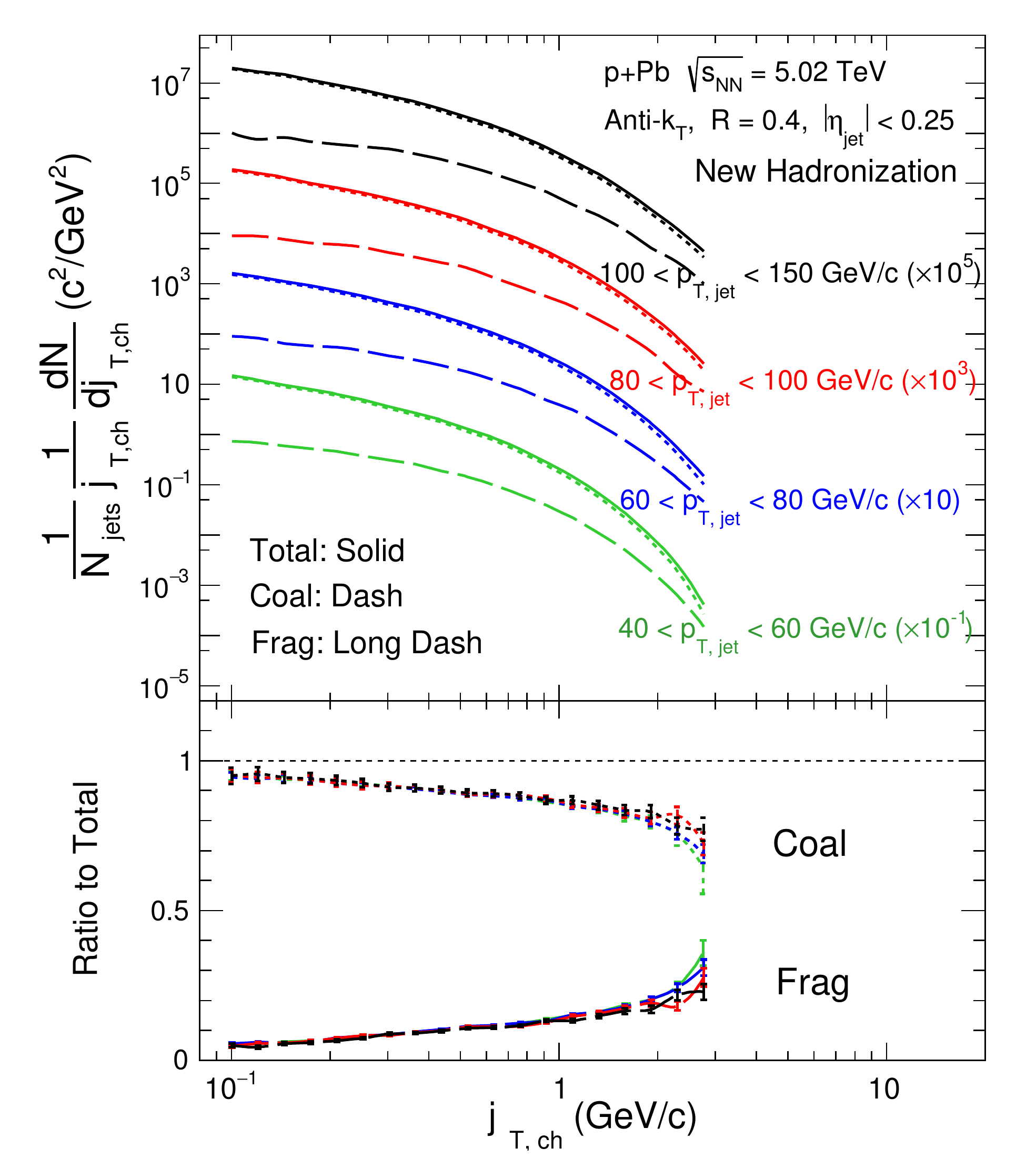}
\caption{Upper panel: The $j_{\rm T}$-dependent jet fragmentation functions of charged particles for four $p_{\rm{T, jet}}$ ranges in p $+$ Pb collisions at $\sqrt{s_{\rm NN}} = 5.02~{\rm TeV}$ from the coalescence and fragmentation parts in the new hadronization mechanism. Bottom panel: The contributions of the coalescence and fragmentation parts to the total $j_{\rm T}$-dependent jet fragmentation function.}
\label{fig.4}
\end{figure}

\begin{figure}[htb]
\centering
\includegraphics
[width=9cm]{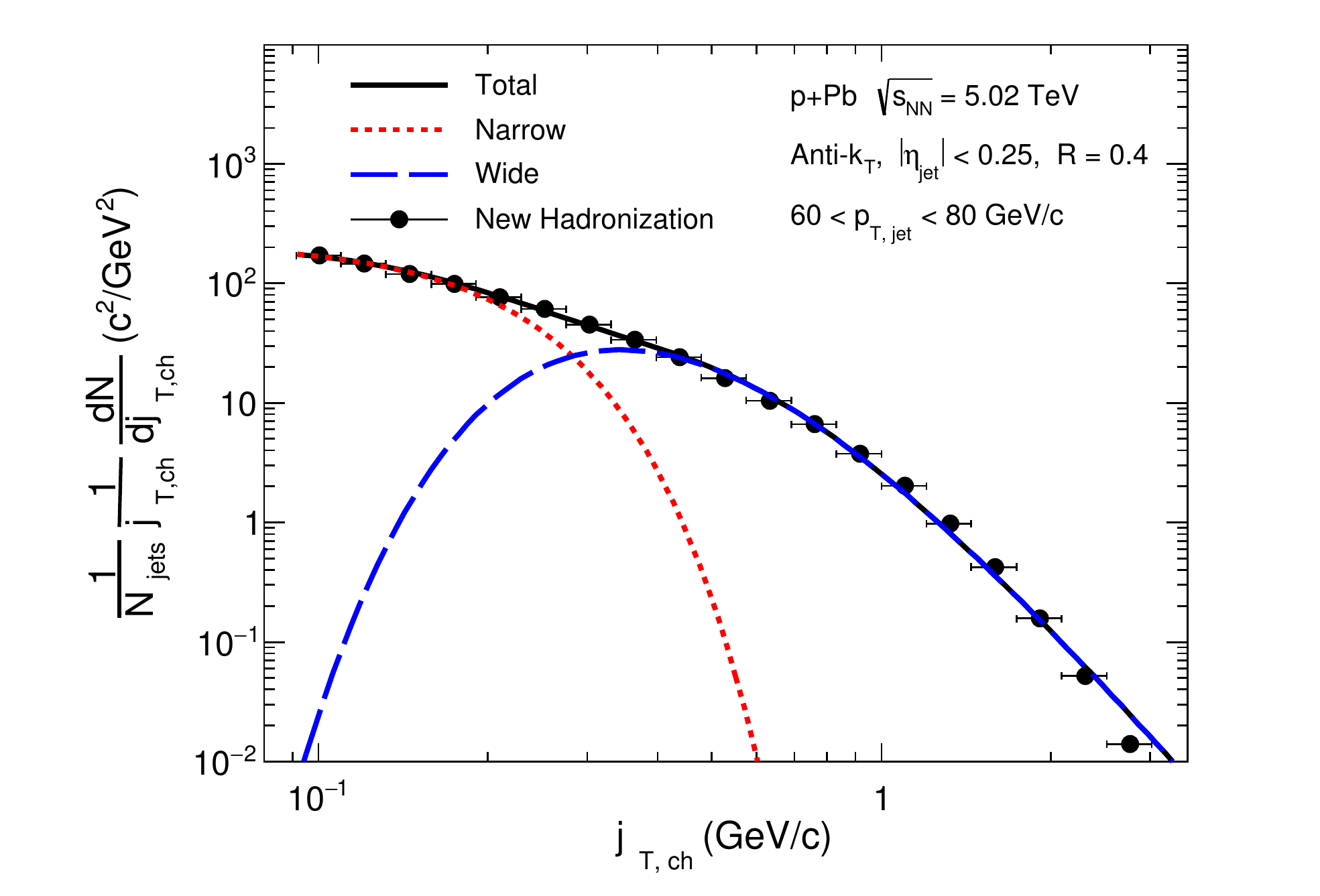}
\caption{The $j_{\rm T}$-dependent jet fragmentation function of charged particles for $60 < p_{\rm{T, jet}} < 80~{{\rm GeV}/c}$ in p $+$ Pb collisions at $\sqrt{s_{\rm NN}} = 5.02~{\rm TeV}$ from the AMPT model with the new hadronization mechanism, which is fitted by Eq.~(\ref{eq.6}) including both narrow and wide parts.}
\label{fig.5}
\end{figure}

\begin{figure}[htb]
\centering
\includegraphics
[width=9cm]{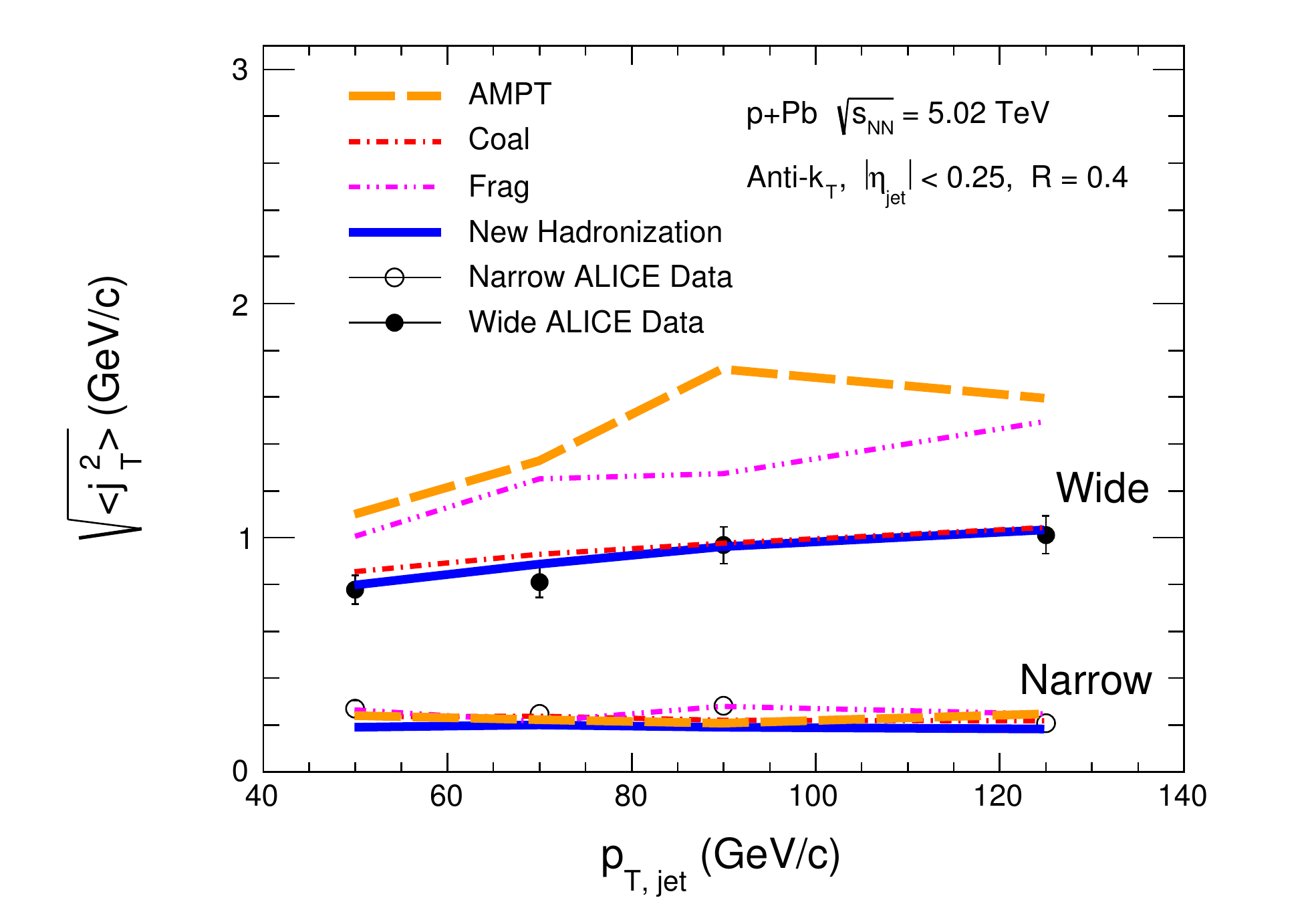}
\caption{The RMS values for the narrow and wide parts for different hadronization mechanisms and their components in p $+$ Pb collisions at $\sqrt{s_{\rm NN}} = 5.02~{\rm TeV}$, compared to the ALICE data~\cite{ALICE:2021JHEP09}. }
\label{fig.6}
\end{figure}

\subsection{The $j_{\rm T}$-dependent JFFs}\label{3.2}

Figure~\ref{fig.2} shows the transverse momentum $j_{\rm T}$-dependent jet fragmentation functions of charged particles inside jets in p $+$ p collisions (top panels) and p $+$ Pb collisions (middle panels) at $\sqrt{s_{\rm NN}} = 5.02~{\rm TeV}$ for four $p_{\rm{T, jet}}$ intervals (between 40 and $150~{\rm GeV}/c$) from the AMPT model with old and new hadronization mechanisms, compared to the ALICE data~\cite{ALICE:2021JHEP09}. In general, more low $j_{\rm T}$ particles are produced by the new hadronization mechanism than that by the old simple coalescence model. Remarkably, the results from the new hadronization mechanism are in good agreement with the experimental data for low and intermediate $j_{\rm T}$ regions. However, it slightly overestimates the experimental data at the high $j_{\rm T}$ region~\footnote{This could be due to the absence of gluons in the AMPT output partons.}. In addition, the bottom panels present the ratios of the JFF in p $+$ Pb to that in p $+$ p collisions for different parton interaction cross sections in ZPC part. Two sets of parton interaction cross sections, $3~\rm mb$ and $0~\rm mb$, are utilized to mimic with and without jet-medium interactions in p $+$ Pb collisions.  We find that the $j_{\rm T}$-dependent JFFs in p $+$ Pb collisions, with different cross sections,  are consistent with that in p $+$ p collisions, almost regardless of parton interaction cross section. It indicates that there is no obvious modification effect from jet-medium interactions or the cold nuclear matter effect on the $j_{\rm T}$-dependent JFFs in p $+$ Pb collisions. 

The upper panel in Fig~\ref{fig.3} shows the $j_{\rm T}$-dependent jet fragmentation functions of charged particles inside jets for four $p_{\rm{T, jet}}$ intervals in p $+$ Pb collisions at $\sqrt{s_{\rm NN}} = 5.02~{\rm TeV}$ from the AMPT model with old and new hadronization mechanisms, compared to the ALICE data. The ratios of the $j_{\rm T}$-dependent JFF for high $p_{\rm{T, jet}}$ intervals with respect to that for the lowest $p_{\rm{T, jet}}$ interval (40 $<p_{\rm{T, jet}}<$ $60~{\rm GeV}$/$c$) are presented in the bottom panel in Fig~\ref{fig.3}. These ratios increase slowly with $j_{\rm T}$ in the low and intermediate $j_{\rm T}$ regions ($j_{\rm T}<1.0$ GeV/c). At $j_{\rm T}>1.0$ GeV/c, the ratios increases rapidly. These features are caused by the fact that the high-$p_{\rm{T, jet}}$ jet includes more particles inside jet cone, especially more large-$j_{\rm T}$ particles, than the low-$p_{\rm{T, jet}}$ jets, which is consistent with the ALICE measurements~\cite{ALICE:2021JHEP09}. The ratio from the new hadronization mechanism is higher than that from the old hadronization mechanism, suggesting that such a ratio could be used as a sensitive probe to study the effect of different hadronization mechanisms on the JFFs.

To further investigate the individual contributions from quark coalescence and string fragmentation to the total JFFs in the new hybrid hadronization mechanism, we compare the $j_{\rm T}$-dependent JFFs of charged particles from quark coalescence, string fragmentation and the total for four $p_{\rm{T, jet}}$ ranges in p $+$ Pb collisions at $\sqrt{s_{\rm NN}} = 5.02~{\rm TeV}$ in upper panel of Fig~\ref{fig.4}. We find that most hadrons inside a jet are produced by the coalescence process. To learn about the exact contribution percentages from two hadronization mechanisms, the ratios of the quark coalescence-contributed JFF to the total JFF and the fragmentation-contributed JFF to the total JFF are shown as functions of $j_{\rm T}$ in the bottom panel of Fig~\ref{fig.4}. We find that $j_{\rm T}$-dependent JFFs are dominantly contributed by the quark coalescence part. The contribution of the string fragmentation part increases with $j_{\rm T}$ and reaches up to 25\% at $j_{\rm T} \sim$ 3 GeV/$c$.

The $j_{\rm T}$-dependent jet fragmentation functions have been fitted with a two-component function in Eq.~(\ref{eq.6}) including a Gaussian function (narrow part or low-$j_{\rm T}$ part) and an inverse gamma function (wide part or high-$j_{\rm T}$ part). Figure~\ref{fig.5} shows our fitting to the $j_{\rm T}$-dependent JFF for $60 < p_{\rm{T, jet}} < 80~{{\rm GeV}/c}$ in p $+$ Pb collisions at $\sqrt{s_{\rm NN}} = 5.02~{\rm TeV}$ from the AMPT model with the new hadronization mechanism. We can see that the two-component function can well describe our results.  This is understandable because the final state radiation is also turned on during our initialization of the jet in HIJING part. Based on the fitting function, we calculate the root-mean-square (RMS) values for narrow and wide parts based on Eqs.~(\ref{eq.7}) and~(\ref{eq.8}). In Fig~\ref{fig.6}, the RMS values from the different hadronization processes and their components are shown as functions of $p_{\rm{T, jet}}$, compared to the ALICE measurements. We find that all RMS values of the narrow part are consistent with the ALICE data, independent of $p_{\rm{T, jet}}$. It indicates that the RMS value of narrow part is not sensitive to hadronization mechanism. However, for the wide part, the AMPT model with the old hadronization mechanism fails and overestimates the ALICE data. On the other hand, the new hadronization mechanism describes well the measured RMS values of wide part which increases with $p_{\rm{T, jet}}$. Note that the experimental data are also described well by the coalescence part in the new hadronization mechanism, which indicates that the $j_{\rm T}$-dependent JFFs are dominantly contributed by the coalescence process. The above comparisons reveal that the RMS value of wide part is more sensitive to different hadronization mechanisms than the narrow part.

\subsection{The $R$ size dependence of $j_{\rm T}$-dependent JFFs}\label{3.3}

\begin{figure}[htb]
\centering
\includegraphics
[width=9cm]{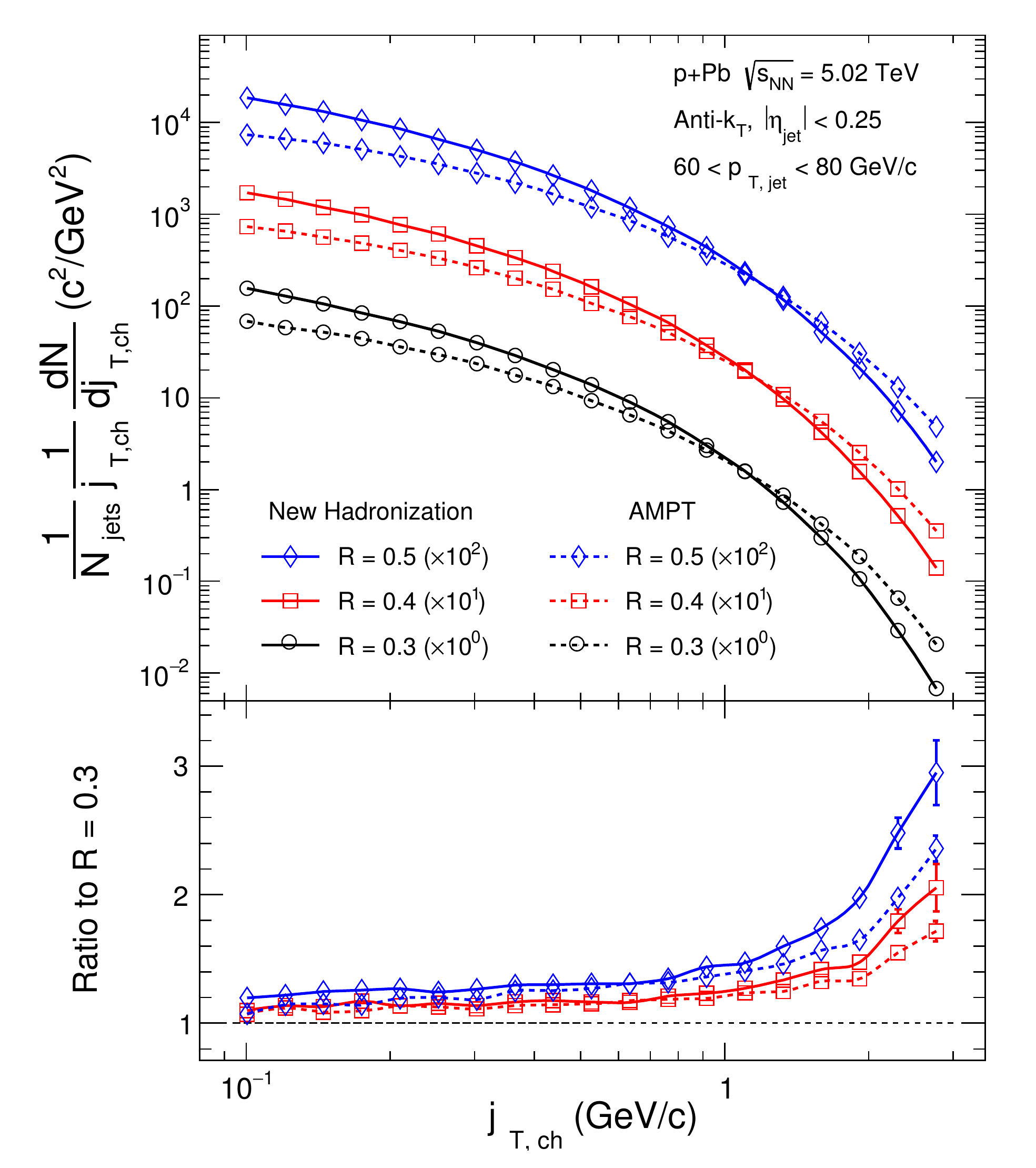}
\caption{Upper panel: The $j_{\rm T}$-dependent jet fragmentation functions of charged particles with three jet radii $R$ = 0.3, 0.4 and 0.5 for $60 < p_{\rm{T, jet}} < 80~{{\rm GeV}/c}$ in p $+$ Pb collisions at $\sqrt{s_{\rm NN}} = 5.02~{\rm TeV}$ from the AMPT model with old (dashed curve) and new hadronization (solid curve) mechanisms. Bottom panel: The ratios of the $j_{\rm T}$-dependent jet fragmentation function for a large jet radius $R$ to that for the smallest jet radius $R$ = 0.3.}
\label{fig.7}
\end{figure}

\begin{figure}[htb]
\centering
\includegraphics
[width=9cm]{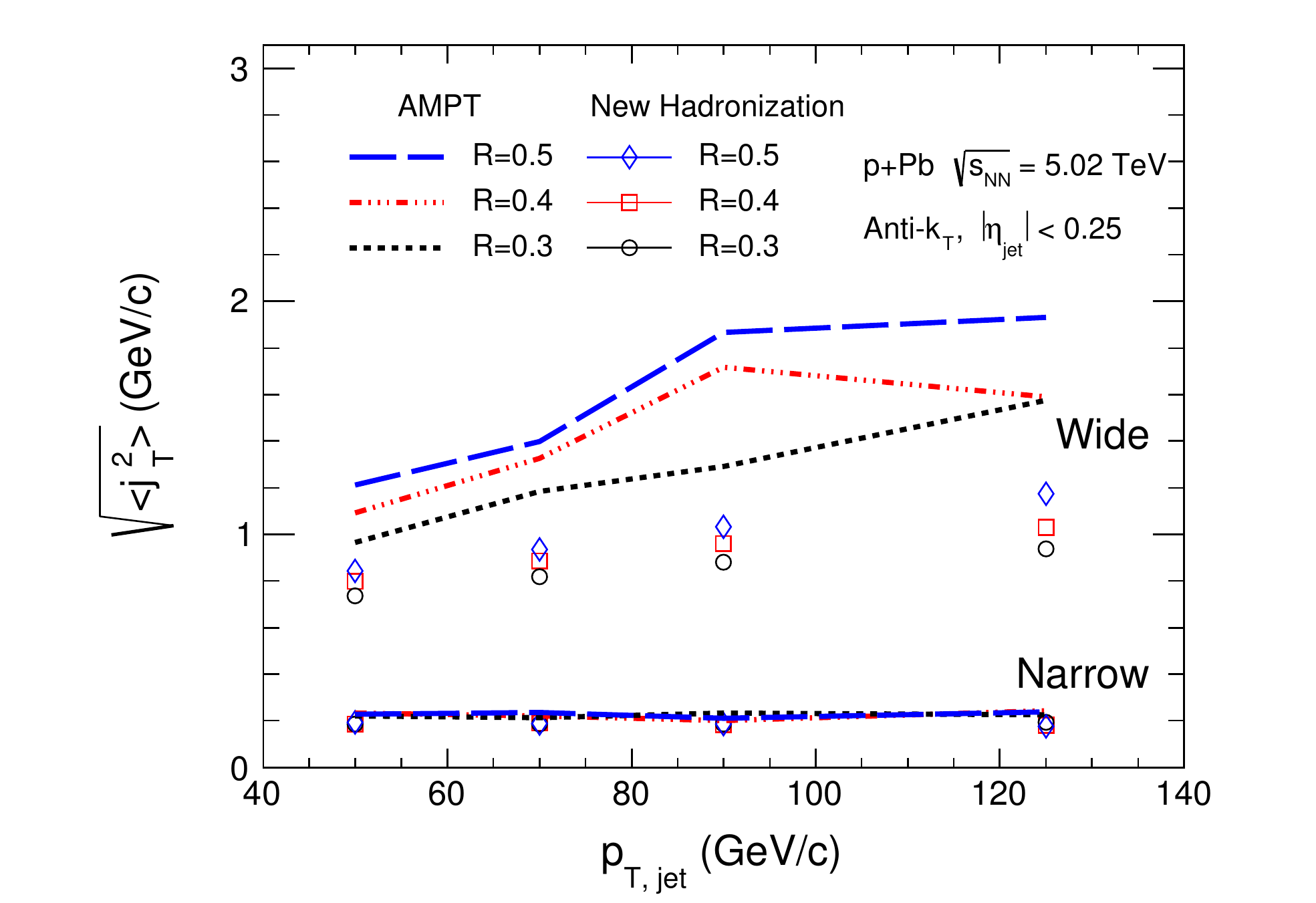}
\caption{The RMS values for the narrow and wide parts for jets with three jet radii $R$ from the AMPT model with old and new hadronization mechanisms in p $+$ Pb collisions at $\sqrt{s_{\rm NN}} = 5.02~{\rm TeV}$.}
\label{fig.8}
\end{figure}

Varying the size of the jet cone can incorporate the particles with different $j_{\rm T}$ into the jet cone.  Three jet cone size radii, $R =$ 0.3, 0.4, and 0.5, are used in our analysis to study the jet cone size dependence of the $j_{\rm T}$-dependent JFFs. The $j_{\rm T}$-dependent JFFs for $60 < p_{\rm{T, jet}} < 80~{{\rm GeV}/c}$ in p $+$ Pb collisions at $\sqrt{s_{\rm NN}} = 5.02~{\rm TeV}$ from the AMPT model with old and new hadronization mechanisms are shown in the upper panel of Fig~\ref{fig.7}. The bottom panel of Fig~\ref{fig.7} shows the ratios of the $j_{\rm T}$-dependent JFF with respect to that in the lowest jet radius $R = 0.3$. We find that the ratio increases slowly with $j_{\rm T}$ in the low and intermediate $j_{\rm T}$ ($j_{\rm T}<$ 1.1 GeV/c) regions, while the ratio increases rapidly with $j_{\rm T}$ in the high $j_{\rm T}$ region, which indicates that the jets with a large jet cone size carry more associated particles, especially for the high-$j_{\rm T}$ particles, compared to the jet with the small jet cone size. On the other hand, we find that the ratio for new hadronization mechanism is larger than that for old hadronization mechanism. The difference is more obvious for a larger jet cone size, which suggests that jets with a larger jet cone size are more sensitive to different hadronization mechanisms.

The corresponding RMS values for different jet cone sizes $R$ as functions of $p_{\rm{T, jet}}$ are shown in Fig~\ref{fig.8}. Note that we do not compare our results with experimental data, since the experimental data for $R =$ 0.3 and 0.5 are unavailable. There is no obvious difference among different jet cone sizes for the RMS value of narrow part, also independent of hadronization mechanism. On the other hand, the RMS value of wide part increases with the jet radius $R$ and jet transverse momentum $p_{\rm{T, jet}}$. However, the AMPT model with the old hadronization mechanism gives larger RMS values of wide part than the new hadronization mechanism. Therefore, we propose that the RMS value of wide part could also be a sensitive probe to study different hadronization mechanisms. 

\begin{figure}[htb]
\centering
\includegraphics
[width=9cm]{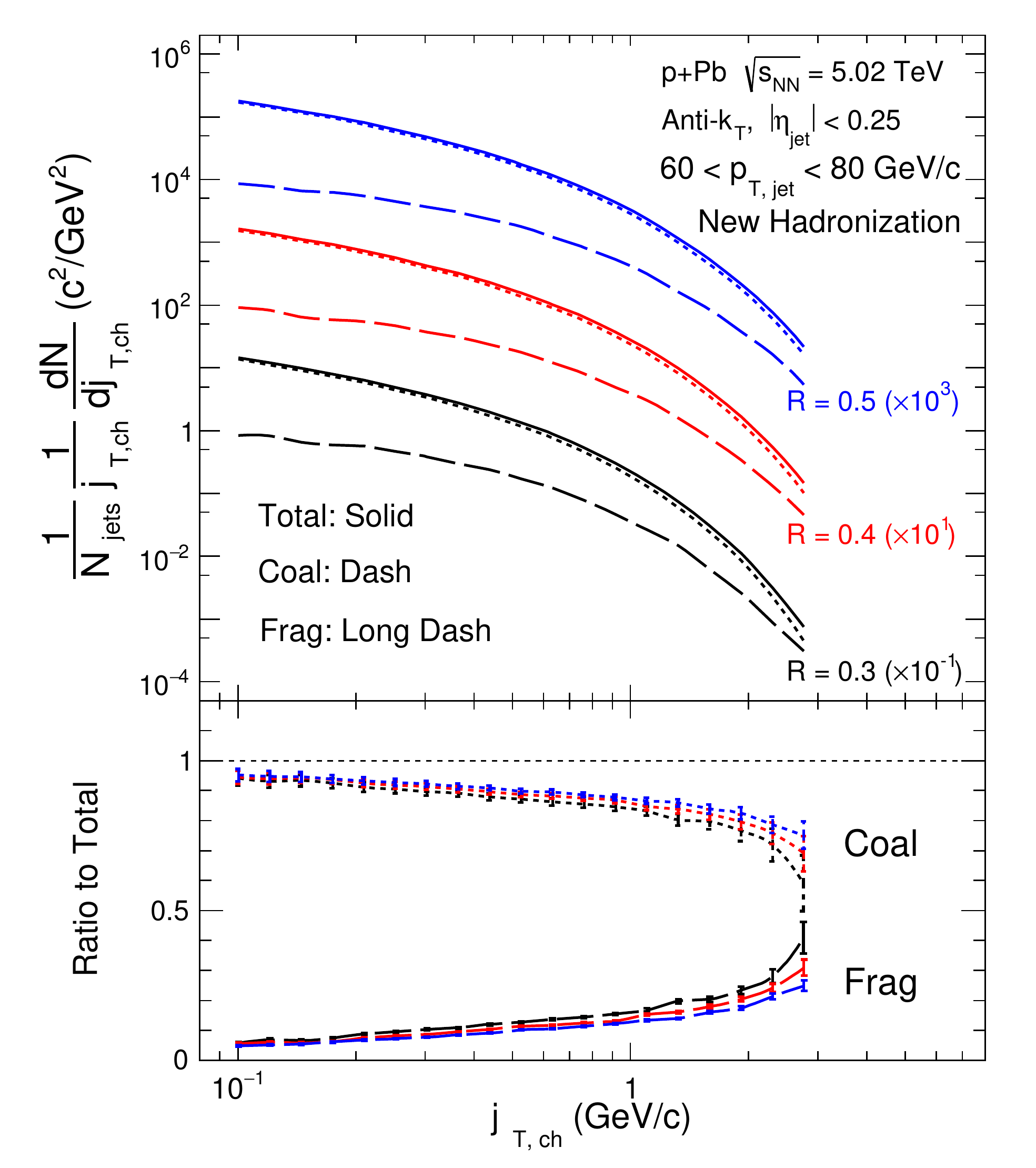}
\caption{Upper panel: The $j_{\rm T}$-dependent jet fragmentation functions of charged particles for three jet cone size radii in p $+$ Pb collisions at $\sqrt{s_{\rm NN}} = 5.02~{\rm TeV}$ from the coalescence and fragmentation parts in the new hadronization mechanism. Bottom panel: The corresponding contributions of the coalescence and fragmentation parts to the total $j_{\rm T}$-dependent jet fragmentation function.}
\label{fig.9}
\end{figure}

For the new hybrid hadronization mechanism, we also compare the individual contributions from quark coalescence and string fragmentation to the total JFFs for different jet cone size radii. The upper panel of Fig.~\ref{fig.9} shows the $j_{\rm T}$-dependent jet fragmentation functions of charged particles for $60 < p_{\rm{T, jet}} < 80~{{\rm GeV}/c}$ with three jet cone size radii ($R =$ 0.3, 0.4, and 0.5) in p $+$ Pb collisions at $\sqrt{s_{\rm NN}} = 5.02~{\rm TeV}$. We observe that $j_{\rm T}$-dependent JFFs are dominantly contributed by the quark coalescence process for any of three jet cone size radii. Quantitatively, the ratios of two hadronization mechanisms to the total JFFs are shown in the bottom panel of Fig.~\ref{fig.9}. We find that the jets with a larger jet cone size include more high-$j_{\rm T}$ particles in the quark coalescence part, which indicates that more particles are produced by the quark coalescence hadronization mechanism due to the large jet cone size. On the other hand, the contribution of the string fragmentation part increases most with $j_{\rm T}$ for the smallest jet cone $R =$ 0.3, which suggests that the string fragmentation mechanism works relatively more efficiently for the smaller jet cone size. In this way, we may be able to study the hadronization mechanism of interest by increasing its contribution to the JFF by using different radii of jets.

\section{Summary}\label{4}

The $j_{\rm T}$-dependent jet fragmentation functions have been investigated in p $+$ p and p $+$ Pb collisions at $\sqrt{s_{\rm NN}} = 5.02~{\rm TeV}$ with a multiphase transport (AMPT) model using old simple quark coalescence mechanism and a new hadronization mechanism which includes both dynamical quark coalescence and string fragmentation schemes. The $j_{\rm T}$-dependent JFFs from the new hadronization mechanism are in better agreement with the ALICE experimental data than the old simple quark coalescence mechanism. Within the new hadronization mechanism framework, the $j_{\rm T}$-dependent JFFs are dominated by the coalescence, while the contribution of the fragmentation increases with $j_{\rm T}$. The ratios of the $j_{\rm T}$-dependent JFFs in p $+$ Pb collisions to those in p $+$ p collisions are consistent with unity, which indicate there are no significant effects of jet quenching or cold nuclear matter in p $+$ Pb collisions. On the other hand, we find that the $j_{\rm T}$-dependent JFFs are sensitive to different hadronization mechanisms. The RMS values of the JFFs for narrow and wide parts have been extracted by a two-component function fitting for different hadronization mechanisms. Unlike the insensitivity for the RMS value of narrow part, the RMS value of wide part increases with the increasing of $p_{\rm{T, jet}}$ and jet cone size $R$. Since the RMS value of wide part is very sensitive to different hadronization mechanisms, we propose that the wide part of JFFs is a good probe to investigate the non-perturbative QCD hadronization effect on jets in small colliding systems. The contribution of the coalescence part is found to increase with the jet cone size $R$, especially for high $j_{\rm T}$. Besides, we would like to emphasize that our definition of JFF is traditional and not infrared-and-collinear safe, which is sensitive to non-perturbative physics such as the hadronization mechanism~\cite{ATLAS:2011EPJC71,Kang:2017glf,Dasgupta:2007wa,Kaufmann:2015PRD92}. On the other hand, it would be more interesting to further investigate whether there is no or less hadronization effect in the recently proposed infrared-and-collinear JFF with the help of subjets~\cite{Caucal:2020JHEP10}. In addition, it would be also interesting to explore more features of $j_{\rm T}$-dependent JFF by using jet grooming techniques~\cite{Larkoski:2014wba,Milhano:2017nzm}. On the other hand, our current work focuses only on small colliding systems, and future studies of $j_{\rm T}$-dependent JFFs in high-energy A $+$ A collisions are expected to explore the effect of hot QCD medium modifications on jet hadronization~\cite{Adolfsson:2020dhm}.

\begin{acknowledgments}
X.-P.D. and G.-L.M. are supported by the National Natural Science Foundation of China under Grants No.12147101, No. 11890714, No. 11835002, No. 11961131011, No. 11421505, the National Key Research and Development Program of China under Contract No. 2022YFA1604900, the Strategic Priority Research Program of Chinese Academy of Sciences under Grant No. XDB34030000, and the Guangdong Major Project of Basic and Applied Basic Research under Grant No. 2020B0301030008. W. Z. is supported by the National Science Foundation (NSF) under Grant No. ACI-2004571 within the framework of the XSCAPE project of the JETSCAPE collaboration.
\end{acknowledgments}

\section*{Appendix}
\label{appen}
\subsection{The $\xi$-dependent JFFs}

The longitudinal momentum of hadrons within a jet, $p_{\|}^{\text {track}}$, is defined as the parallel component of the momentum of a constituent particle with respect to the reconstructed jet axis,
\begin{equation}\label{eq.9}
p_{\|}^{\text {track}}=\frac{\vec{p}_{\text {jet}} \cdot \vec{p}_{\text {track}}}{\left|\vec{p}_{\text {jet}}\right|},
\end{equation}
where $\vec{p}_{\text {jet}}$ is the momentum of the jet and $\vec{p}_{\text {track}}$ is the momentum of the tracks.
The longitudinal momentum fraction of the tracks ($z_{\rm h}$) with respect to the jet and variable $\xi$ are calculated by,
\begin{equation}\label{eq.10}
z_{\rm h}=\frac{p_{\|}^{\text {track}}}{p_{\text {jet}}}, \quad \xi=\log \frac{1}{z_{\rm h}}.
\end{equation}
Meanwhile, the $\xi$-dependent jet fragmentation functions are defined by
\begin{equation}\label{eq.11}
D(\xi) = \frac{1}{N_{\text {jets}}} \frac{\rm{d} N_{\text {track}}} {\rm d\xi},
\end{equation}
where $N_{\text {jets}}$ is the number of the reconstructed jets and $N_{\text {track}}$ is the number of the tracks inside the reconstructed jet cone.

For the $\xi$-dependent jet fragmentation function measurement, we use the jet reconstruction and background subtraction methods as in the analysis of $j_{\rm T}$-dependent jet fragmentation function. We apply the same kinematic cuts as the CMS experiment~\cite{CMS:2015CMS15}, which requires the transverse momenta of particles $p_{\rm T} > 0.5~\rm{GeV}/c$ and the jet pseudorapidity $\left|\eta_{\rm jet}\right| < 1.5$.

\begin{figure*}[!htbp]
\centering
\includegraphics
[width=18cm]{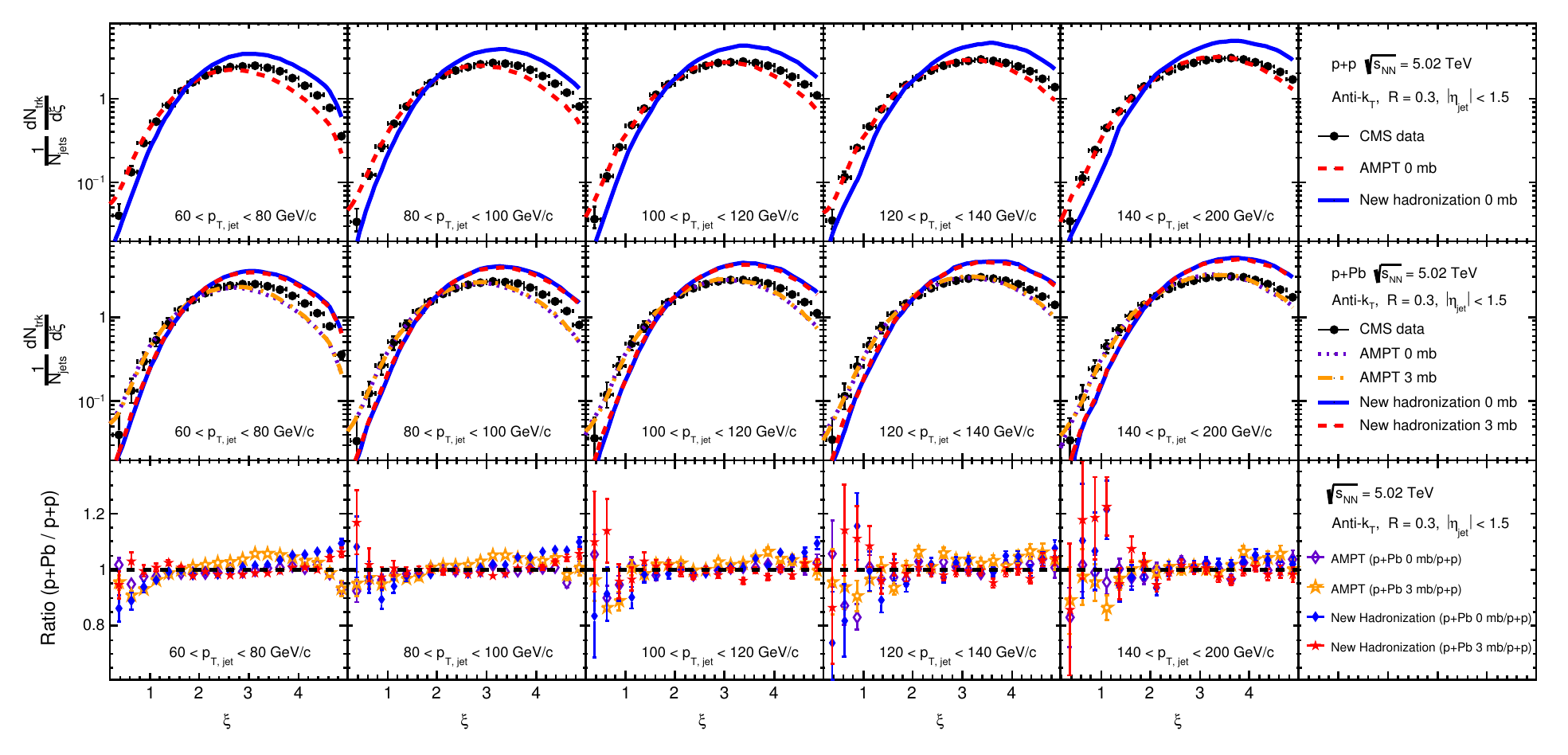}
\caption{Top panels: The $\xi$-dependent jet fragmentation functions of tracks inside jets for five $p_{\rm{T, jet}}$ ranges in p $+$ p collisions at $\sqrt{s_{\rm NN}} = 5.02~{\rm TeV}$ from the AMPT model with old and new hadronization mechanisms, compared to the CMS data~\cite{CMS:2015CMS15}. Middle panels: Same as the top panels but for p $+$ Pb collisions at $\sqrt{s_{\rm NN}} = 5.02~{\rm TeV}$. Bottom panels: The ratios of the $\xi$-dependent jet fragmentation function in p $+$ Pb collisions (0 or 3 mb) to that in p $+$ p collisions (0 mb) for the AMPT model with the old and new hadronization mechanisms.}
\label{fig.10}
\end{figure*}

Figure~\ref{fig.10} presents the longitudinal momentum $\xi$-dependent jet fragmentation functions of tracks inside jets in p $+$ p collisions (top panels) and p $+$ Pb collisions (middle panels) at $\sqrt{s_{\rm NN}} = 5.02~{\rm TeV}$ for five bins of $60 < p_{\rm{T, jet}} < 200~{{\rm GeV}/c}$ from the AMPT model with old and new hadronization mechanisms with the same set of parameters for our $j_{\rm T}$ calculations. Compared to the CMS data~\cite{CMS:2015CMS15}, we find that the old hadronization model can better describe the experimental data than the new hadronization model, which is the opposite to the $j_{\rm T}$ performance. This is partially because we did not adjust any parameters in the other AMPT parts for our specific study of jet fragmentation functions. Similarly, it shows that ratios of the $D(\xi)$ also do not present any evidence for jet-medium interaction or the cold nuclear matter effects in p $+$ Pb collisions.

\begin{figure*}[!htbp]
\centering
\includegraphics
[width=18cm]{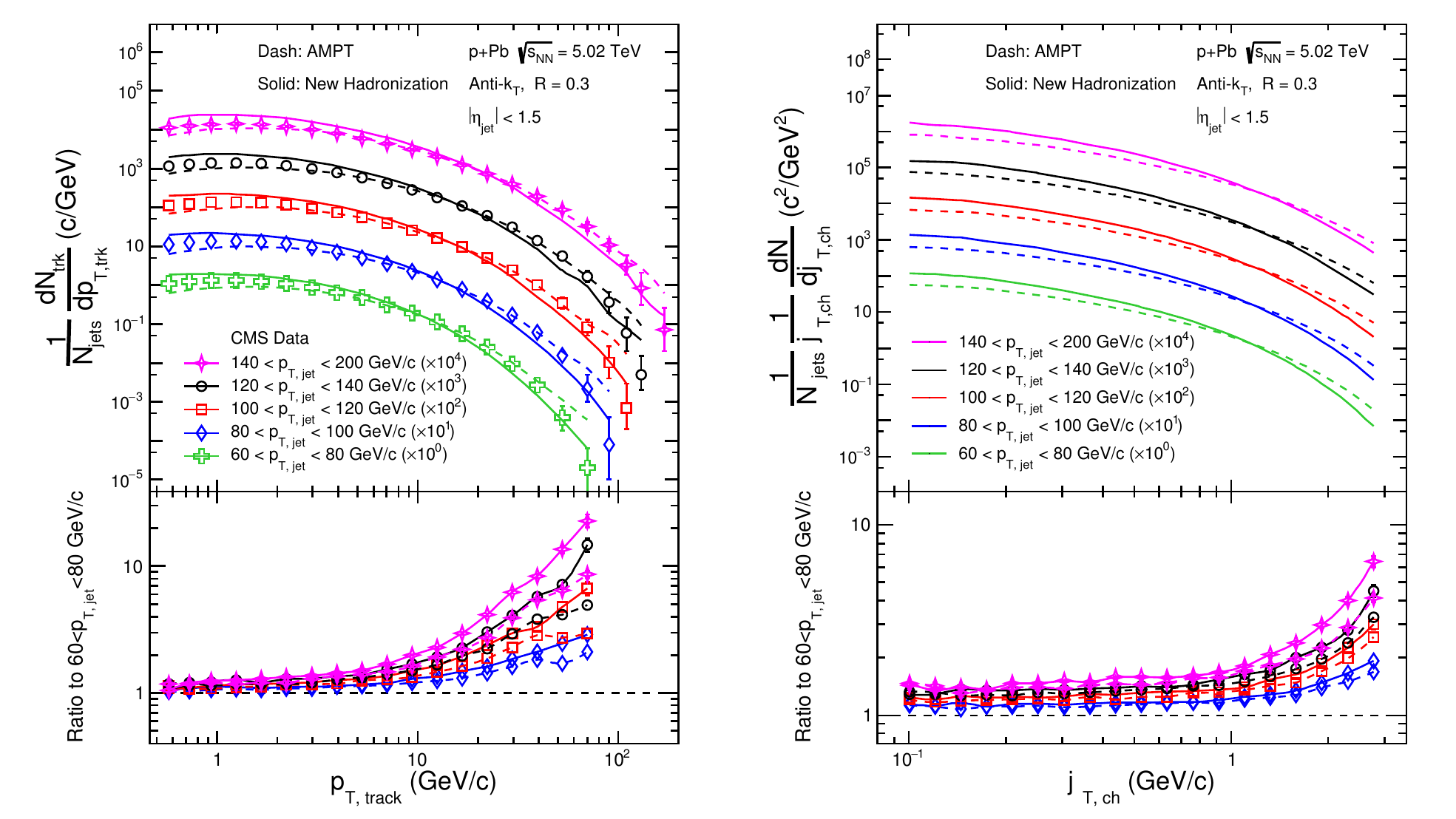}
\caption{Left panel: The $p_{\rm T}$ spectrum of tracks inside the jet cones for five $p_{\rm{T, jet}}$ ranges in p $+$ Pb collisions at $\sqrt{s_{\rm NN}} = 5.02~{\rm TeV}$ from the AMPT model with old (dashed curve) and new hadronization mechanisms (solid curve), compared to the CMS data~\cite{CMS:2015CMS15}. Right panel: Same as the left panel but for the $j_{\rm T}$ distributions.}
\label{fig.11}
\end{figure*}

\begin{figure*}[!htbp]
\centering
\includegraphics
[width=18cm]{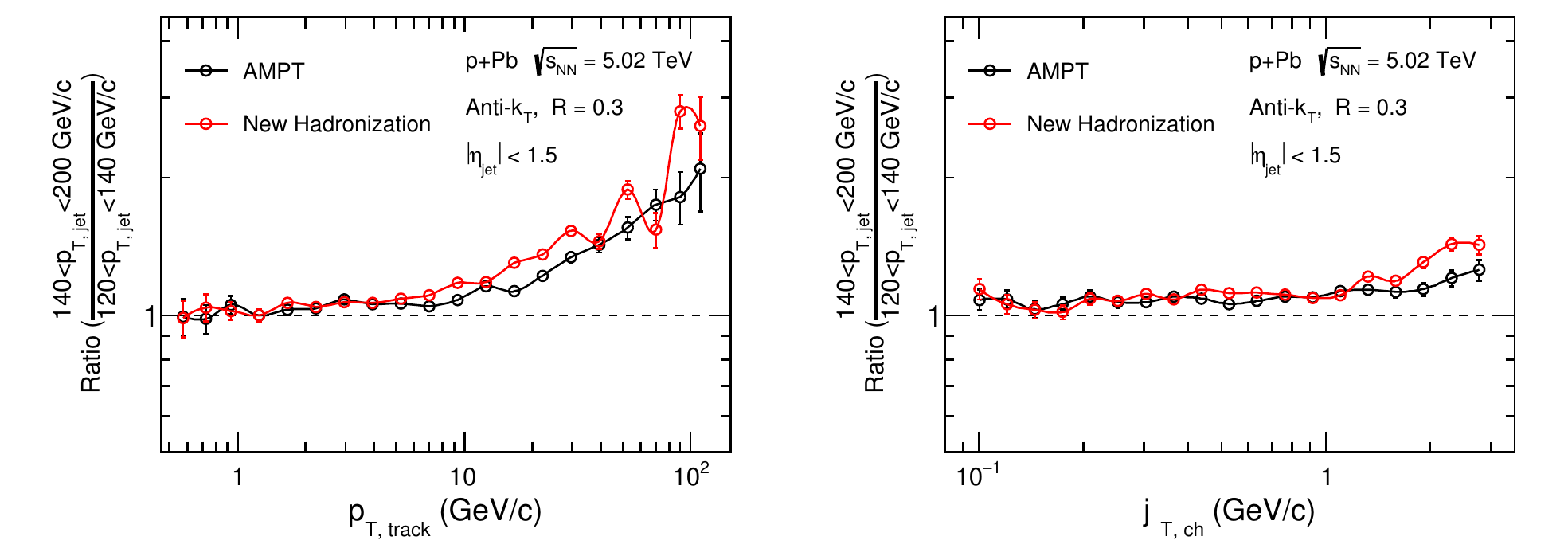}
\caption{Left panel: The ratios of the $p_{\rm T}$ spectrum of tracks inside the jet cone between the two highest $p_{\rm{T, jet}}$ ranges in p $+$ Pb collisions at $\sqrt{s_{\rm NN}} = 5.02~{\rm TeV}$ from the AMPT model with old and new hadronization mechanisms. Right panel: Same as the left panel but for the $j_{\rm T}$ distributions.}
\label{fig.12}
\end{figure*}

\subsection{The $p_{\rm T}$ spectrum inside jet cones}

The left panel in Fig.~\ref{fig.11} shows the $p_{\rm T}$ spectrum of tracks inside the jet cone in p $+$ Pb collisions at $\sqrt{s_{\rm NN}} = 5.02~{\rm TeV}$ for $60 < p_{\rm{T, jet}} < 200~{{\rm GeV}/c}$ from the AMPT model with old and new hadronization mechanisms, compared to the CMS data~\cite{CMS:2015CMS15}. We find that the CMS measurements for $p_{\rm T}$ spectrum of tracks inside the jet cones can be better described by the old hadronization model. Similarly, we see more low $p_{\rm T}$ particles in the jet cone from the new hadronization model than those from the old hadronization model. In addition, we also calculate the jet fragmentation transverse momentum ($j_{\rm T}$) distributions using the same experimental condition for the $p_{\rm T}$ spectrum measurements, which is shown in the right panel in Fig.~\ref{fig.11}. The two hadronization mechanisms show the similar behaviors for the ratios of the $p_{\rm T}$ spectrum and the $j_{\rm T}$ distributions between the higher $p_{\rm{T, jet}}$ interval with respect to that in $60 < p_{\rm{T, jet}} < 80~{{\rm GeV}/c}$.

Moreover, we further compare the ratio between the two highest $p_{\rm{T, jet}}$ intervals ($140 < p_{\rm{T, jet}} < 200~{{\rm GeV}/c}$ over $120 < p_{\rm{T, jet}} < 140~{{\rm GeV}/c}$), since it is supposed that higher $p_{\rm{T, jet}}$ interval results should contain weaker non-perturbation effects. Figure~\ref{fig.12} shows that, as expected, the two hadronization mechanisms present almost the same behavior, which indicates that the very high-$p_{\rm{T, jet}}$ jets results are dominated by the perturbative effect and the contributions from non-perturbation processes are significantly suppressed.

\textbf{Data Availability Statement:}
This manuscript has associated data in a data repository. [Authors’ comment: The datasets used and/or analyzed during the current study are available from the corresponding author on reasonable request.]

\bibliography{references.bib}

\end{document}